\documentclass[conference]{IEEEtran}
\usepackage{amsmath}
\usepackage{amssymb}
\usepackage[dvips]{graphicx}
\usepackage{epsfig}

\newcommand{\X}{\mathbf{X}}

\newcommand{\z}{\mathbf{z}}
\newcommand{\h}{\mathbf{h}}

\newcommand{\hh}{\mathbf{H}}
\newcommand{\w}{\mathbf{w}}

\newcommand{\I}{\mathbf{I}}

\newtheorem{lemma:bitenergylow}{Lemma}
\newtheorem{lemma:bitenergyhigh}[lemma:bitenergylow]{Lemma}

\newtheorem{prop:asympcap}{Theorem}
\newtheorem{prop:flashminbitenergy}[prop:asympcap]{Theorem}
\newtheorem{prop:flashbitenergy}[prop:asympcap]{Theorem}
\newtheorem{prop:pasympcap}[prop:asympcap]{Theorem}

\begin{document}


\title{Relay Beamforming Strategies for Physical-Layer Security}



%
\author{\authorblockN{Junwei Zhang and Mustafa Cenk Gursoy}
\authorblockA{Department of Electrical Engineering\\
University of Nebraska-Lincoln, Lincoln, NE 68588\\ Email:
junwei.zhang@huskers.unl.edu, gursoy@engr.unl.edu}}

\maketitle
\begin{abstract}\footnote{This work was supported by the National Science Foundation under Grant CCF -- 0546384 (CAREER).}
In this paper, collaborative use of relays to form a beamforming
system  and provide physical-layer security is investigated. In
particular, amplify-and-forward (AF) relay beamforming designs under
total and individual relay power constraints are studied with the
goal of maximizing the secrecy rates when perfect channel state
information (CSI) is available. In the AF scheme, not having
analytical solutions for the optimal beamforming design under both
total and individual power constraints, an iterative algorithm is
proposed to numerically obtain the optimal beamforming structure and
maximize the secrecy rates. Robust beamforming designs in the
presence of imperfect CSI are investigated for decode-and-forward
(DF) based relay beamforming, and optimization frameworks are
provided.

\end{abstract}

\section{introduction}
The broadcast nature of wireless transmissions allows for the
signals to be received by all users within the communication range,
making wireless communications vulnerable to eavesdropping. The
problem of secure transmission in the presence of an eavesdropper
was first studied from an information-theoretic perspective in
\cite{wyner} where Wyner considered a wiretap channel model. Wyner
showed that secure communication is possible without sharing a
secret key if the eavesdropper's channel is a degraded version of
the main channel, and identified the rate-equivocation region and
established the secrecy capacity of the degraded discrete memoryless
wiretap channel. The secrecy capacity is defined as the maximum
achievable rate from the transmitter to the legitimate receiver,
which can be attained  while keeping the eavesdropper completely
ignorant of the transmitted messages. Later, Wyner's result was
extended to the Gaussian channel in \cite{cheong} and recently to
fading channels in \cite{Liang} and \cite{Gopala}. In addition to
the single antenna case, secrecy in multi-antenna models is
addressed in \cite{shafiee}
 and \cite{khisti}.  Regarding multiuser models, Liu  \emph{et al.} \cite{Liu} presented inner and
outer bounds on secrecy capacity regions for broadcast and
interference channels. The secrecy capacity of  the multi-antenna
broadcast channel is obtained in \cite{Liu1}. Additionally, it is
well known that even if they are equipped with single-antennas
individually, users can cooperate to form a distributed
multi-antenna system by performing relaying. When channel side
information (CSI) is exploited, relay nodes can collaboratively work
similarly as in a MIMO system to build a virtual beam towards the
receiver. Relay beamforming research has attracted much interest
recently (see e.g., \cite{Jing}--\cite{Gan1} and references
therein). Cooperative relaying under secrecy constraints was also
recently studied in \cite{dong}--\cite{jzhang} . In \cite{dong}, a
decode-and-forward (DF) based cooperative protocol is considered,
and a beamforming system is designed for secrecy capacity
maximization or transmit power minimization. For amplify-and-forward
(AF), suboptimal closed-form solutions  that optimize bounds on
secrecy capacity are proposed in \cite{dong1}. However, in those
studies, the analysis is conducted only under total relay power
constraints and perfect CSI. In our recent work\cite{jzhang}, we
studied the problem of DF beamforming under both total and individual
power constraints. It is shown that the total power constraint leads
to a closed-form solution. The design under individual relay power
constraints is formulated as an optimization problem which is shown
to be easily solved using two different approaches, namely
semidefinite programming and second-order cone programming. In this
paper, we study amplify-and-forward (AF) relay beamforming under
both total and individual power constraints. We also extend our previous study by considering two robust beamforming design methods for DF relaying under imperfect CSI.

\section{Channel Model}

\begin{figure}
\begin{center}
\includegraphics[width = 0.5\textwidth]{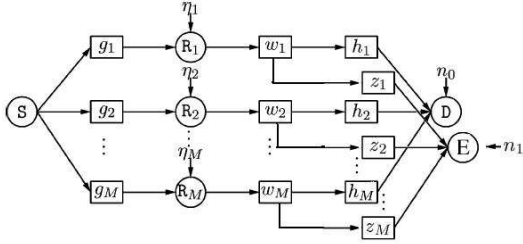}
\caption{Channel Model} \label{fig:channel}
\end{center}
\end{figure}

We consider a communication channel with a source $S$, a destination
$D$, an eavesdropper $E$, and $M$ relays $\{R_m\}_{m=1}^M$ as depicted
in Figure.\ref{fig:channel}. We assume that there is no direct link
between $S$ and $D$, and $S$ and $E$. We also assume that relays work synchronously
by multiplying the signals to be transmitted with complex weights $\{w_m\}$ to produce a virtual beam
point to destination and eavesdropper. We denote the channel fading coefficient
between $S$ and $R_m$ as $g_m\in \mathbb{C}$, the fading coefficient between
$R_m$ and $D$ as $h_m\in \mathbb{C}$, and the fading coefficient between $R_m$
and $E$ as $z_m\in \mathbb{C}$. In this model, the source $S$ tries to transmit
confidential messages to $D$ with the help of the relays  while keeping the
eavesdropper $E$ ignorant of the information.
It's obvious that our channel is a two-hop relay network. In the
first hop, the source $S$ transmits $x_s$ to relays with power
$E[|x_s|^2]=P_s$. The received signal at $R_m$ is given by
\begin{align}
y_{r,m}=g_m x_s+\eta_m
\end{align}
where $\eta_m$ is the background noise that has a Gaussian
distribution with zero mean and variance of $N_m$. In first part of
the paper, we focus on amplify-and-forward relaying. Robust
design for DF will be discussed subsequently.

In the AF scenario, the received
signal at the $m^{th}$ relay $R_m$ is directly multiplied by
$l_mw_m$ without decoding, and forwarded to $D$. The relay output
can be written as
\begin{align}
x_{r,m}=w_m l_m (g_m x_s+ \eta_m).
\end{align}
The scaling factor,
\begin{align}
l_m=\frac{1}{\sqrt{|g_m|^2P_s+N_m}},
\end{align}
is used to ensure $E[|x_{r,m}|^2]=|w_m|^2$. There are two kinds of
power constraint for relays. First one is  a total relay power
constraint in the following form: $||\w||^2=\w^\dagger \w\leq P_T$
where $\w=[w_1,...w_M]^T$, $P_T$ is the maximum total power for all
relays. $(\cdot)^T$ and $(\cdot)^\dagger $ denote the transpose
and conjugate transpose, respectively, of a matrix or vector. In a
multiuser network such as the relay system we study in this paper,
it is practically more relevant to consider individual power
constraints as wireless nodes generally operate under such
limitations. Motivated by this, we can impose $|w_m|^2 \leq p_m
\forall m $ or equivalently $|\w|^2 \leq \mathbf{p}$ where
$|\cdot|^2$ denotes the element-wise norm-square operation and
$\mathbf{p}$ is a column vector that contains the components
$\{p_m\}$. $p_m$ is the maximum power for the $m^{th}$ relay node.

The
received signals at the destination $D$ and eavesdropper $E$ are the
superposition of the messages sent by the relays. These received
signals are expressed, respectively, as
\begin{align}
y_d&=\sum_{m=1}^M h_m w_m l_m (g_m x_s +\eta_m) +n_0, ~\text{and} \\
 y_e&=\sum_{m=1}^M z_m w_m l_m (g_m x_s+\eta_m) +n_1.
\end{align}
where $n_0$ and $n_1$ are the Gaussian background noise components
at $D$ and $E$, respectively, with zero mean and variance $N_0$.
Now, it is easy to compute  the received SNR at $D$ and $E$ as
\begin{align}
\Gamma_d&=\frac{|\sum_{m=1}^M h_m g_m l_m w_m|^2 P_s}{\sum_{m=1}^M |h_m|^2l_m^2 |w_m|^2 N_m +N_0}, ~\text{and} \\
\Gamma_e&=\frac{|\sum_{m=1}^M z_m g_m l_m w_m|^2 P_s}{\sum_{m=1}^M
|z_m|^2l_m^2 |w_m|^2 N_m +N_0}.
\end{align}
The secrecy rate is now given by
\begin{align}
R_s&=I(x_s;y_d)-I(x_s;y_e)\\
&=\log(1+\Gamma_d)-\log(1+\Gamma_e)\\
&=\log\Bigg(\frac{\sum_{m=1}^M |z_m|^2l_m^2 |w_m|^2 N_m
+N_0}{\sum_{m=1}^M |h_m|^2l_m^2 |w_m|^2 N_m +N_0}\times \nonumber \\
&\frac{|\sum_{m=1}^M h_m g_m l_m w_m|^2 P_s+\sum_{m=1}^M
|h_m|^2l_m^2 |w_m|^2 N_m +N_0}{|\sum_{m=1}^M z_m g_m l_m w_m|^2
P_s+\sum_{m=1}^M |z_m|^2l_m^2 |w_m|^2 N_m +N_0}\Bigg)\label{srate}.
\end{align}
where $I(\cdot;\cdot)$ denotes the mutual information. In this
paper, we address the joint optimization of $\{w_m\}$ with the aid
perfect CSI and hence identify the optimum collaborative relay
beamforming (CRB) direction that maximizes the secrecy rate
in (\ref{srate}).

\section{Optimal Beamforming for AF case}
 Let us define
\begin{align}
\mathbf{h_g}&=[h_1^* g_1^* l_1,... ,h_M^* g_M^* l_M]^T, \\
\mathbf{h_z}&=[z_1^* g_1^* l_1,... ,z_M^* g_M^* l_M]^T, \\
\mathbf{D_h}&=\text{Diag}(|h_1|^2l_1^2N_1,...,|h_M|^2l_M^2N_M),  ~\text{and}\\
\mathbf{D_z}&=\text{Diag}(|z_1|^2l_1^2N_1,...,|z_M|^2l_M^2N_M).
\end{align}
where superscript $*$ denotes conjugate operation. Then, the
received SNR at the destination and eavesdropper can be
reformulated, respectively, as
\begin{align}
\Gamma_d&=\frac{ P_s  \w^\dagger \mathbf{h_g} \mathbf{h_g}^\dagger
\w }{\w^\dagger \mathbf{D_h} \w  +N_0}
         =\frac{ P_s  tr (\mathbf{h_g} \mathbf{h_g}^\dagger \w \w^\dagger) }{tr( \mathbf{D_h} \w\w^\dagger)
         +N_0}, ~\text{and}\\
\Gamma_e&=\frac{P_s  \w^\dagger \mathbf{h_z}\mathbf{h_z}^\dagger
\w}{\w^\dagger \mathbf{D_z} \w  +N_0} =\frac{ P_s  tr (\mathbf{h_z}
\mathbf{h_z}^\dagger \w \w^\dagger) }{tr( \mathbf{D_z} \w\w^\dagger)
+N_0}
\end{align}
where $tr(\cdot)$ represent the trace of a matrix. It is obvious
that we only have to maximize the term inside the logarithm
function in (\ref{srate}). With these notations, we can write the objective function
of the optimization problem as
\begin{align}
&\frac{1+\Gamma_d}{1+\Gamma_e}\nonumber \\&=\frac{1+\frac{ P_s
\w^\dagger \mathbf{h_g} \mathbf{h_g}^\dagger \w }{\w^\dagger
\mathbf{D_h} \w +N_0}}{1+\frac{P_s  \w^\dagger
\mathbf{h_z}\mathbf{h_z}^\dagger \w}{\w^\dagger \mathbf{D_z} \w
+N_0}}\\
&=\frac{\w^\dagger \mathbf{D_h} \w  +N_0+P_s  \w^\dagger
\mathbf{h_g} \mathbf{h_g}^\dagger \w }{\w^\dagger \mathbf{D_z} \w
+N_0+P_s  \w^\dagger \mathbf{h_z}\mathbf{h_z}^\dagger \w} \times
\frac{\w^\dagger \mathbf{D_z} \w +N_0}{\w^\dagger \mathbf{D_h} \w
+N_0}\\
&=\frac{N_0+tr((\mathbf{D_h}+P_s\mathbf{h_g} \mathbf{h_g}^\dagger)\w
\w^\dagger)}{N_0+tr((\mathbf{D_z}+P_s\mathbf{h_z}
\mathbf{h_z}^\dagger)\w \w^\dagger)} \times
\frac{N_0+tr(\mathbf{D_z} \w \w^\dagger)}{N_0+tr(\mathbf{D_h} \w
\w^\dagger)}.
\end{align}
If we denote $t_1=\frac{N_0+tr((\mathbf{D_h}+P_s\mathbf{h_g}
\mathbf{h_g}^\dagger)\w
\w^\dagger)}{N_0+tr((\mathbf{D_z}+P_s\mathbf{h_z}
\mathbf{h_z}^\dagger)\w \w^\dagger)}$,
$t_2=\frac{N_0+tr(\mathbf{D_z} \w \w^\dagger)}{N_0+tr(\mathbf{D_h}
\w \w^\dagger)}$, and  define $\X\ \triangleq \w \w ^\dagger$ using
the similar semidefinite programming method as
described in \cite{jzhang}, we can express the optimization problem as
\begin{align}
\begin{split}
&\max_{\X, t_1, t_2} ~~~t_1t_2  \\
&\text{s.t} ~~ \text{tr}\left(\X\left(\mathbf{D_z}-
t_2\mathbf{D_h}\right)\right)\geq N_0(t_2-1)\\
&~~ \text{tr}\left(\X\left(\mathbf{D_h}+P_s\mathbf{h_g}
\mathbf{h_g}^\dagger- t_1\left(\mathbf{D_z}+P_s\mathbf{h_z}
\mathbf{h_z}^\dagger\right)\right)\right)\geq N_0(t_1-1)\\
&\text{and} ~~\text{diag}(\X)\leq \mathbf{p},~~(\text{and/or}~~
\text{tr}(\X) \leq P_T)~~~ \text{and} ~~~\X \succeq 0.
\end{split}
\end{align}
where  $\X\succeq 0$ means that $\X$  is a symmetric positive
semi-definite matrix. Since $\X$ by definition is a rank one
matrix, finding the optimal weights is in general a nonconvex optimization problem. Thus, we above
ignore the rank constraint, and hence employ semidefinite relaxation
(SDR) \cite{luo}.  If the matrix $\X_{opt}$ obtained by solving the above
optimization problem happens to be rank one, then its principal
component will be the optimal solution to the original problem.

Notice that this formulation is applied to both total relay power
constraint and individual relay power constraint which are
represented by $tr(\X) \leq P_T$ and $diag(\X)\leq \mathbf{p}$,
respectively. When there is only total power constraint, we can
easily compute the maximum values of $t_1$ and $t_2$ separately
since now we have Rayleigh quotient problems.
\begin{align}
t_{1,u}&=\max_{\w^\dagger \w\leq P_T}\frac{\w^\dagger \mathbf{D_h}
\w  +N_0+P_s  \w^\dagger \mathbf{h_g} \mathbf{h_g}^\dagger \w
}{\w^\dagger \mathbf{D_z} \w +N_0+P_s  \w^\dagger
\mathbf{h_z}\mathbf{h_z}^\dagger
\w}\\
&=\max_{\w^\dagger \w\leq P_T}\frac{\w^\dagger (\mathbf{D_h}
+\frac{N_0}{P_T}+P_s\mathbf{h_g} \mathbf{h_g}^\dagger )\w
}{\w^\dagger (\mathbf{D_z} \w +\frac{N_0}{P_T}+P_s
\mathbf{h_z}\mathbf{h_z}^\dagger
)\w} \\
&=\lambda_{max}\left(\mathbf{D_h}+\frac{N_0}{P_T}\I+P_s \mathbf{h_g}
\mathbf{h_g}^\dagger,\mathbf{D_z}+\frac{N_0}{P_T}\I+P_s\mathbf{h_z}
\mathbf{h_z}^\dagger\right).
\end{align}
where $\lambda_{\max}(\mathbf{A},\mathbf{B})$ is the largest
generalized eigenvalue of the matrix pair $(\mathbf{A},\mathbf{B})$
\footnote{For a Hermitian matrix $\mathbf{A} \in \mathbb{C}^{n\times
n}$ and positive definite matrix $\mathbf{B} \in \mathbb{C}^{n\times
n}$, $(\lambda,\psi)$ is referred to as a generalized eigenvalue --
eigenvector pair of $(\mathbf{A},\mathbf{B})$ if $(\lambda,\psi)$
satisfy $\mathbf{A} \psi=\lambda \mathbf{B} \psi$ \cite{matrix}.}.

Similarly, maximum values of $t_2$ under total power constraint is
\begin{align}
t_{2,u}&=\max_{\w^\dagger \w\leq P_T} \frac{\w^\dagger \mathbf{D_z}
\w +N_0}{\w^\dagger \mathbf{D_h} \w+N_0} \\
&=\max_{\w^\dagger \w\leq P_T}\frac{\w^\dagger( \mathbf{D_z}+
\frac{N_0}{P_T})\w}{\w^\dagger (\mathbf{D_h} +\frac{N_0}{P_T})\w}\\
&=\lambda_{max}\left(\mathbf{D_z}+\frac{N_0}{P_T}\I,\mathbf{D_h}+\frac{N_0}{P_T}\I\right).
\end{align}
When there are individual power constraints imposed on the relays,
The maximum values $t_{1,i,u}$ and $t_{2,i,u}$ \footnote{Subscripts
$i$ in $t_{1,i,u}$ and $t_{2,i,u}$ are used to denote that these are
the maximum values in the presence of individual power constraints.}
for $t_1$ and $ t_2$ is obtained by solving following optimization
problem:
\begin{align}
\begin{split}\label{t1}
&\max_{\X, t_1} ~~~t_1   \\
&\text{s.t}~~ \text{tr}\left(\X\left(\mathbf{D_h}+P_s\mathbf{h_g}
\mathbf{h_g}^\dagger- t_1\left(\mathbf{D_z}+P_s\mathbf{h_z}
\mathbf{h_z}^\dagger\right)\right)\right)\geq N_0(t_1-1)\\
&\text{and} ~~\text{diag}(\X)\leq \mathbf{p},~~~ \text{and} ~~~\X
\succeq 0,
\end{split}
\intertext{and}
\begin{split}\label{t2}
&\max_{\X, t_2} ~~~t_2  \\
&\text{s.t} ~~ \text{tr}\left(\X\left(\mathbf{D_z}-
t_2\mathbf{D_h}\right)\right)\geq N_0(t_2-1)\\
&\text{and} ~~\text{diag}(\X)\leq \mathbf{p},~~~ \text{and} ~~~\X
\succeq 0.
\end{split}
\end{align}
In fact, for any value of $t_1$, the feasible set in (\ref{t1}) is
convex. If, for any given $t_1$, the convex feasibility problem
\begin{align}
\begin{split}\label{t11}
&\text{find} ~~\X  \\
&\text{s.t}~~ \text{tr}\left(\X\left(\mathbf{D_h}+P_s\mathbf{h_g}
\mathbf{h_g}^\dagger- t_1\left(\mathbf{D_z}+P_s\mathbf{h_z}
\mathbf{h_z}^\dagger\right)\right)\right)\geq N_0(t_1-1)\\
&\text{and} ~~\text{diag}(\X)\leq \mathbf{p},~~~ \text{and} ~~~\X
\succeq 0,
\end{split}
\end{align}
is feasible, then we have $t_{1,i,u}\geq t_1$. Conversely, if the
convex feasibility optimization problem (\ref{t11}) is not feasible,
then we conclude $t_{1,i,u} <t_1$. Therefore, we can check whether
the optimal value $t_{1,i,u}$ of the quasiconvex optimization
problem in (\ref{t1}) is smaller than or greater than a given value
$t_1$ by solving the convex feasibility problem (\ref{t11}). If the
convex feasibility problem (\ref{t11}) is feasible then we know
$t_{1,i,u}\geq t$. If the convex feasibility problem (\ref{t11}) is
infeasible, then we know that  $t_{1,i,u}< t_1$. Based on this
observation, we can use a simple  bisection algorithm to solve the
quasiconvex optimization problem (\ref{t1}) by solving a convex
feasibility problem (\ref{t11}) at each step. We assume that the
problem is feasible, and start with an interval $[l ,u]$ known to
contain the optimal value $t_{1,i,u}$. We then solve the convex
feasibility problem at its midpoint $t_1 = (l + u)/2$ to determine
whether the optimal value is larger or smaller than $t$. We update
the interval accordingly to obtain a new interval. That is, if $t_1$
is feasible, then we set $l = t_1$, otherwise, we choose $u = t_1$
and solve the convex feasibility problem  again. This procedure is
repeated until the width of the interval is smaller than the given
threshold. Then, we conclude that $t_{1,i,u}=l$.

Similarly, $t_{2,i,u}$ can be obtained with the same bisection
algorithm  by repeatedly solving the following feasibility problems:
\begin{align}
\begin{split}\label{t22}
&\text{find} ~~\X  \\
&\text{s.t} ~~ \text{tr}\left(\X\left(\mathbf{D_z}-
t_2\mathbf{D_h}\right)\right)\geq N_0(t_2-1)\\
&\text{and} ~~\text{diag}(\X)\leq \mathbf{p},~~~ \text{and} ~~~\X
\succeq 0.
\end{split}
\end{align}
To solve the convex feasibility problem, one can use the
well-studied interior-point based methods as well. We use the
well-developed interior point method based package SeDuMi
\cite{sedumi}, which produces a feasibility certificate if the
problem is feasible, and its popular interface  Yalmip
\cite{yalmip}.

Note that for both total and individual power constraints,  the
maximum values of $t_1$ and $t_2$ are obtained separately above, and
these values are in general attained by different $\X = \w
\w^\dagger$. Now, the following strategy can be used to obtain
achievable secrecy rates. For those $\X$ values that correspond to
$t_{1,i,u} $ and $t_{1,u}$ (i.e., the maximum $t_1$ values under
individual and total power constraints, respectively), we can
compute the corresponding $t_2=\frac{N_0+tr(\mathbf{D_z} \w
\w^\dagger)}{N_0+tr(\mathbf{D_h} \w \w^\dagger)}$  and denote them
as $t_{2,i,l}$ and $t_{2,l}$ for individual and total power
constraints, respectively. Then, $\log (t_{1,i,u}t_{2,i,l})$ and
$\log (t_{1,u}$$t_{2,l})$ will serve as our amplify-and-forward
achievable rates for individual and total power constraints,
respectively. With the achievable rates, we propose the following
algorithm to iteratively search over $t_1$ and $ t_2$ to get the
optimal $t_{1,o}$ and $t_{2,o}$ that maximize the product $t_1 t_2$
by checking the following feasibility problem.
\begin {align}\label{feasible1}
\begin{split}
& \text{find} ~~~~\X\succeq 0  \\
&\text{s.t} ~  \text{tr}\left(\X\left(\mathbf{D_z}-
t_2\mathbf{D_h}\right)\right)\geq N_0(t_2-1)\\
& \text{tr}\left(\X\left(\mathbf{D_h}+P_s\mathbf{h_g}
\mathbf{h_g}^\dagger- t_1\left(\mathbf{D_z}+P_s\mathbf{h_z}
\mathbf{h_z}^\dagger\right)\right)\right)\geq N_0(t_1-1)\\
&\text{and}   ~~\text{tr}(\X) \leq P_T~~~\text{if there is  total power constraint}, \\
 &\text{or}~~\text{diag}(\X)\leq \mathbf{p} ~~\text{if there is  individual power constraint}.
\end{split}
\end{align}
\subsection{Proposed Algorithm}
Define the resolution $\Delta t=\frac{t_{1,u}}{N}$ or $\Delta t=\frac{t_{1,i,u}}{N}$ for some  large $N$ for total and individual power constraints, respectively. \\
\begin{enumerate}
\item  Initialize $t_{1,o}=t_{1,u}$ , $t_{2,o}=t_{2,l}$
when total power constraint is imposed,  and $t_{1,o}=t_{1,i,u}$,
$t_{2,o}=t_{2,i,l}$ when individual power constraint is imposed.
Initialize the iteration index $i=N$.
\item Set $t_1=i\Delta t$.   If $t_1t_{2,u}<t_{1,o}t_{2,o}$
(total power constraint) or $t_1t_{2,i,u}<t_{1,o}t_{2,o}$
(individual power constraint), then go to Step (3). Otherwise,
\begin{enumerate}
\item  Let  $ t_2=\frac{t_{1,o}t_{2,o}}{t_1}$. Check the
feasibility problem (\ref{feasible1}). If it is infeasible,  go to
step (3). If it is feasible, use the bisection algorithm in
(\ref{feasible1}) with $t_1$ to get the maximum possible values of
$t_2$ and denote this maximum as $t_{2,m}$. The initial interval in
the above bisection algorithm can be chosen as
$[\frac{t_{1,o}t_{2,o}}{t_1}, t_{2,u}]$ or
$[\frac{t_{1,o}t_{2,o}}{t_1}, t_{2,i,u}]$ depending on the  power
constraints.
\item  Update $t_{1,o}=t_1$, $t_{2,o}=t_{2,m}$ ,
$i=i-1$. Go back to step (2).
\end{enumerate}
\item  Solve the following convex problem to get the optimal $\X$
\begin{align}
\begin{split}
& \min_{\X} ~~~~\text{tr}(\X)  \\
&\text{s.t} ~  \text{tr}\left(\X\left(\mathbf{D_z}-
t_{2,o}\mathbf{D_h}\right)\right)\geq N_0(t_{2,o}-1)\\
&~~~\text{tr}\left(\X\left(\mathbf{D_h}+P_s\mathbf{h_g}
\mathbf{h_g}^\dagger- t_{1,o}\left(\mathbf{D_z}+P_s\mathbf{h_z}
\mathbf{h_z}^\dagger\right)\right)\right)\\
&~~~\geq N_0(t_{1,o}-1)\\
&~~\X\succeq 0 \,\,\,\text{and}
\\
&~~\text{tr}(\X) \leq P_T~~~\text{if there is  total power constraint}, \\
 &~~\text{diag}(\X)\leq \mathbf{p} ~~\text{if there is  individual power constraint}.
\end{split}
\end{align}
\end{enumerate}
\section{Robust Beamforming Design for DF case}\label{sec:robust}
In the second hop, we can also employ decode and forward
transmission scheme. In this scheme, each relay $R_m$ first decodes
the message $x_s$ and normalizes it as $ x_s'=x_s/\sqrt{P_s}$.
Subsequently, the normalized message is multiplied by the weight
factor $w_m$ to generate the transmitted signal $x_r=w_m x_s'$. The
output power of each relay $R_m$ is given by
\begin{align}
E[|x_r|^2]=E[|w_m  x_s'|^2]=|w_m|^2.
\end{align}
The received signals at the destination $D$ and eavesdropper $E$ are
the superpositions of the signals transmitted from the relays. These
signals can be expressed, respectively, as
\begin{align}
y_d&=\sum_{m=1}^M h_m w_m  x_s' +n_0=\h^\dagger \w x_s' +n_0, \quad \text{and} \\
y_e&=\sum_{m=1}^M z_m w_m  x_s' +n_1 =\mathbf{z}^\dagger \w x_s'
+n_1.
\end{align}
Additionally, we have defined $\mathbf{h}=[h_1^*,....h_M^*]^T,
\mathbf{z}=[z_1^*,....z_M^*]^T$, The metrics of interest are the
received SNR levels at $D$ and $E$, which are given by
\begin{align}
\Gamma_d&=\frac{|\sum_{m=1}^M h_m w_m|^2}{N_0} \,\, \text{and} \,\,
\Gamma_e&=\frac{|\sum_{m=1}^M z_m w_m|^2}{N_0}.
\end{align}
It has been proved that the secrecy rate $R_s$ over the channel
between the relays and destination is
\begin{align}
R_s&=I(x_s;y_d)-I(x_s;y_e)\\
&=\log(1+\Gamma_d)-\log(1+\Gamma_e)\\
&=\log\left(\frac{N_0+|\sum_{m=1}^M h_m w_m|^2}{N_0+|\sum_{m=1}^M
z_m w_m|^2}\right). \label{eq:secrecyrate}
\end{align}
By using the semidefinite relaxation (SDR) approach to approximate
the problem as a convex semidefinite programming (SDP) problem, the
beamforming design for decode-and-forward under perfect CSI with
individual power constraint can be formed as the following optimization problem \cite{jzhang}:
\begin{align}
\begin{split}
&\max_{\X, t} ~~~t \label{SDR2} \\
&\text{s.t} ~~ \text{tr}(\X(\h\h^\dagger- t\z\z^\dagger))\geq N_0(t-1),\\
&\text{and} ~~\text{diag}(\X)\leq \mathbf{p}, ~~~\text{ and} ~~~\X
\succeq 0.
\end{split}
\end{align}
which can be solved efficiently by interior point methods with a
bisection algorithm \cite{jzhang}.

Systems robust against channel mismatches can be obtained by two approaches. In
most of the robust beamforming methods, the perturbation is modeled as a
deterministic one with bounded norm which leads to a worst case
optimization. The other approach applied to the case in which the
CSI error is unbounded is the statistical approach which provides
the robustness in the form of confidence level measured by
probability.

We define $\hat{\hh}=\hat{\h}\hat{\h}^\dagger$ and
$\hat{\mathbf{Z}}=\hat{\z}\hat{\z}^\dagger$ as the channel
estimators, and $\tilde{\hh}=\hh-\hat{\hh}$ and
$\tilde{\mathbf{Z}}=\mathbf{Z}-\hat{\mathbf{Z}}$ as the estimation
errors. First, consider the worst case optimization. In the worst
case assumption, $\tilde{\hh}$ and $\tilde{\mathbf{Z}}$ are bounded
in their Frobenius norm as $||\tilde{\hh}|| \leq \epsilon_H$,
$||\tilde{\mathbf{Z}}|| \leq \epsilon_Z$, where $\epsilon_H,
\epsilon_Z $ are assumed to be upper bounds on the channel
uncertainty. Based on the result of \cite{Bengtsson}, the robust
counterpart of the previously discussed optimization problem can be
written as
\begin{align}
\begin{split}
&\max_{\X, t} ~~~t \label{DFworst} \\
&\text{s.t} ~~ \text{tr}(\X((\hat{\hh}-\epsilon_H\I)- t(\hat{\mathbf{Z}}+\epsilon_Z \I))\geq N_0(t-1),\\
&\text{and} ~~\text{diag}(\X)\leq \mathbf{p}, ~~~ \text{and} ~~~\X
\succeq 0.
\end{split}
\end{align}
Note that the total power constraint $tr(\X)\leq P_T$ can be added
into the formulation or substituted for the individual power
constraint in (\ref{DFworst}). This problem can be solved the same
way as discussed in \cite{jzhang}.

However, the worst-case approach requires the norms to be bounded,
which is usually not satisfied in practice. Also, this approach is
too pessimistic since the probability of the worst-case may be
extremely low. Hence,  statistical approach is a good alternative in
certain scenarios. In our case, we require the probability of the
non-outage for secrecy transmission is greater than the predefined
threshold $\varepsilon$ by imposing
\begin{gather}
Pr\left( \frac{N_0 + \text{tr}((\hat{\hh}+\tilde{\hh})\X)}{N_0 +
\text{tr}((\hat{\mathbf{Z}}+\tilde{\mathbf{Z}})\X)} \geq t\right)
\nonumber
\\= Pr\left(\text{tr}\left(\X(\hat{\hh}+\tilde{\hh}-
t(\hat{\mathbf{Z}}+\tilde{\mathbf{Z}}))\geq
N_0(t-1)\right)\right)\geq \varepsilon.
\end{gather}
Now, the optimization problem under imperfect CSI can be expressed
as
\begin{align}
\begin{split}
&\max_{\X, t} ~~~t \label{robust df} \\
&\text{s.t} ~~ Pr\left(\text{tr}\left(\X(\hat{\hh}+\tilde{\hh}- t(\hat{\mathbf{Z}}+\tilde{\mathbf{Z}}))\geq N_0(t-1)\right)\right)\geq \varepsilon,\\
&\text{and} ~~\text{diag}(\X)\leq \mathbf{p} \,\,( \text{or}
~~\text{tr}(\X)\leq P_T), ~~~ \text{and} ~~~\X \succeq 0.
\end{split}
\end{align}
If relays are under individual power constraints, we use
$diag(\X)\leq \mathbf{p}$. Otherwise, for the case of total power
constraint, we use $tr(\X)\leq P_T$. We can also impose both
constraints in the optimization. For simplicity of the analysis we
assume that the components of the Hermitian channel estimation error
matrices $\tilde{\hh}$ and $\tilde{\mathbf{Z}}$ are independent,
zero-mean, circularly symmetric, complex Gaussian random variables
with variances $\sigma_{\tilde{H}}^2$ and $\sigma_{\tilde{Z}}^2$.
Now, we can rearrange the probability in the constraint as
\begin{align}
Pr\left(\text{tr}\left((\hat{\hh}-t\hat{\mathbf{Z}}+\tilde{\hh}-
t\tilde{\mathbf{Z}})\X\right)\geq (t-1)N_0)\right).
\end{align}
Let us define
$y=\text{tr}\left((\hat{\hh}-t\hat{\mathbf{Z}}+\tilde{\hh}-
t\tilde{\mathbf{Z}}) \X \right)$. For given $\X$, $\hat{\hh}$, and
$\hat{\mathbf{Z}}$, we know from the results of \cite{Chalise} that
$y$ is a Gaussian distributed random variable with mean
$\mu=\text{tr}\left((\hat{\hh}-t\hat{\mathbf{Z}})\X\right)$ and
variance
$\sigma_y^2=(\sigma_{\tilde{H}}^2+t^2\sigma_{\tilde{Z}}^2)\,\text{tr}(\X\X^\dagger)$.
Then, the non-outage probability  can be written as
\begin{align}
Pr(y\geq (t-1)N_0) &=\int_{ (t-1)N_0}^\infty
\frac{1}{\sqrt{2\pi}\sigma_y}\exp\left(-\frac{(y-\mu)^2}{2\sigma_y^2}\right)\\
&=\frac{1}{2}-\frac{1}{2}\text{erf}\left(\frac{(t-1)N_0-\mu}{\sqrt{2}\sigma_y}\right)\geq
\varepsilon,
\end{align}
or equivalently as,
\begin{align}
\frac{(t-1)N_0-\mu}{\sqrt{2}\sigma_y}\leq
\text{erf}^{-1}(-2\varepsilon+1).
\end{align}
Note that $\varepsilon$ should be close to one for good performance.
Thus, both $-2\varepsilon+1$ and
$\frac{(t-1)N_0-\mu}{\sqrt{2}\sigma_y}$ should be negative valued.
Note further that we have $tr(\X\X^\dagger)=\|\X\|^2$, and hence
$\sigma_y=\sqrt{\sigma_{\tilde{H}}^2+t^2\sigma_{\tilde{Z}}^2}\|\X\|$.
Then, this constraint can be written as
\begin{align}
\|\X\| \leq \frac{(t-1)N_0-\mu}{\sqrt{2
(\sigma_{\tilde{H}}^2+t^2\sigma_{\tilde{Z}}^2)}\text{erf}^{-1}(-2\varepsilon+1)}.
\end{align}
As a result, the optimization problem becomes
\begin{align}
\begin{split}
&\max_{\X, t} ~~~t \label{robust df1} \\
&\text{s.t} ~~||\X|| \leq \frac{(t-1)N_0-\mu}{\sqrt{2
(\sigma_{\tilde{H}}^2+t^2\sigma_{\tilde{Z}}^2)}\text{erf}^{-1}(-2\varepsilon+1)},\\
&\text{and} ~~\text{diag}(\X)\leq \mathbf{p} ( \text{or}
~~\text{tr}(\X)\leq P_T), ~~~ \text{and} ~~~\X \succeq 0.
\end{split}
\end{align}
Using the same bisection search, we can solve this optimization
numerically.

\section{Numerical Results}\label{sec:simulation}
We assume  that $\{g_m\}$, $\{h_m\}, \{z_m\}$ are complex,
circularly symmetric Gaussian random variables with zero mean and
variances $\sigma_g^2$, $\sigma_h^2$, and $\sigma_z^2$ respectively.
Moreover, each figure is plotted for fixed realizations of the
Gaussian channel coefficients. Hence, the secrecy rates in the plots
are instantaneous secrecy rates

\begin{figure}
\begin{center}
\includegraphics[width = 0.45\textwidth]{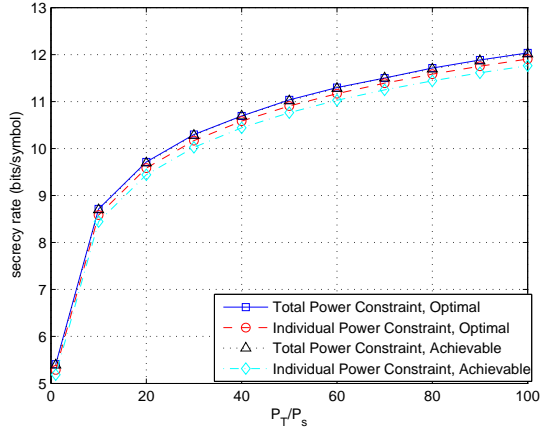}
\caption{AF secrecy rate vs. $P_T/P_s$. $ \sigma_g= 10, \sigma_h=2,
\sigma_z=2, M=10$. } \label{fig:AF1}
\end{center}
\end{figure}

In Fig. \ref{fig:AF1}, we plot the secrecy rate for
amplify-and-forward collaborative relay beamforming system for both
individual and total power constraints. We also provide the result
of suboptimal achievable secrecy rate for comparison. The fixed
parameters are $\sigma_g= 10, \sigma_h=2, \sigma_z=2$,
$N_m=1$,$N_0=1$ and $M=10$. Since the AF secrecy rates depend on
both the source and relay powers, the rate curves are plotted as a
function of $P_T/P_s$. We assume that the relays have equal powers in
the case in which individual power constraints are imposed, i.e.,
$p_i = P_T/M$. It is immediately seen from the figure that the
achievable rates for both total and individual power constraints are
very close to the corresponding optimal ones. Thus, the achievable
beamforming scheme is a good alternative in the amplify-and-forward
relaying case due to the fact that it has much less computational
burden. Moreover, we interestingly observe that imposing individual
relay power constraints leads to only small losses in the secrecy
rates with respect to the case in which we have total relay power
constraints.

\begin{figure}
\begin{center}
\includegraphics[width = 0.45\textwidth]{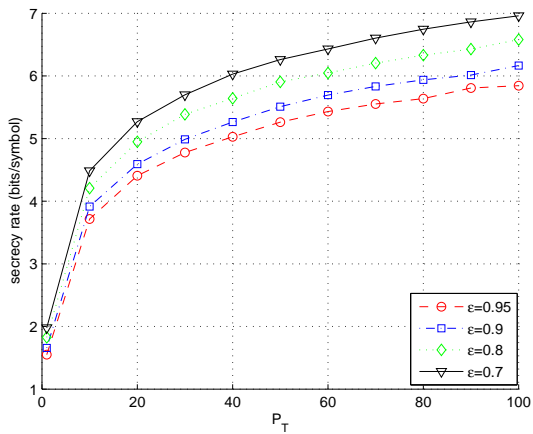}
\caption{DF second secrecy rate vs. $P_T$ under different
$\varepsilon$.  } \label{fig:DFroubst}
\end{center}
\end{figure}

In Fig. \ref{fig:DFroubst},  we plot the maximum second hop secrecy
rate of decode-and-forward that we can achieve for different power
$P_T$ and non-outage probability $\varepsilon$ values. In this
simulation, we fix $M=5$. $\hat{\h}$ and $\hat{\z}$ are randomly
picked from Rayleigh fading with $\sigma_{\hat{h}}=1$ and
$\sigma_{\hat{z}}=2$, and we assume that estimation errors are
inversely proportional to $P_T$. More specifically, in our
simulation, we have $\sigma_{\tilde{H}}^2=0.1/P_T$ and
$\sigma_{\tilde{Z}}^2=0.2/P_T$. We also assume the relays are
operating under equal individual power constraints, i.e.,
$p_i=\frac{P_T}{M}$. It is immediately observed in Fig.
\ref{fig:DFroubst} that smaller rates are supported under higher
non-outage probability requirements. In particular, this figure
illustrates that our formulation and the proposed optimization
framework can be used to determine how much secrecy rate can be
supported at what percentage of the time. For instance, at $P_T =
100$, we see that approximately 7 bits/symbol secrecy rate can be
attained 70 percent of the time (i.e., $\varepsilon = 0.7$) while
supported secrecy rate drops to about 5.8 bits/symbol when
$\varepsilon = 0.95$.

\end{document}